\documentclass[twocolumn,appendixfloats,tighten]{aastex6}
\usepackage{newtxtext,newtxmath}
\usepackage[T1]{fontenc}
\usepackage{ae,aecompl}

\usepackage{amsmath}	
\usepackage{amssymb}	

\usepackage{ctable}
\usepackage{url}
\usepackage{xspace}
\usepackage[normalem]{ulem}
\usepackage{hyperref}
\usepackage[all]{hypcap}
\usepackage{graphicx}
\usepackage{color}

\def\msun{{\rm\,M_\odot}}

\def\msun{{\rm\,M_\odot}} 
\def\lsun{{\rm\,L_\odot}}

\newcommand{\be}{\begin{equation}}
\newcommand{\ee}{\end{equation}}

\def\h2{${\rm\,H_2}$}

\def\gsim{ \lower .75ex \hbox{$\sim$} \llap{\raise .27ex \hbox{$>$}} }
\def\lsim{ \lower .75ex \hbox{$\sim$} \llap{\raise .27ex \hbox{$<$}} }

\newcommand{\beq}{\begin{equation}}
\newcommand{\eeq}{\end{equation}}

\usepackage{color}

\usepackage{array}
\newcolumntype{H}{>{\setbox0=\hbox\bgroup}c<{\egroup}@{}}

\begin{document}

\title{An upper limit on primordial magnetic fields from ultra-faint dwarf galaxies}
\author{Mohammadtaher Safarzadeh\altaffilmark{1,2} and Abraham Loeb\altaffilmark{1} }

\altaffiltext{1}{ Harvard-Smithsonian Center for Astrophysics, 60 Garden St. Cambridge, MA, USA, \href{mailto:msafarzadeh@cfa.harvard.edu}{msafarzadeh@cfa.harvard.edu}}
\altaffiltext{2}{School of Earth and Space Exploration, Arizona State University, AZ, USA}

\begin{abstract}
The presence of primordial magnetic fields increases the minimum halo mass in which star formation is possible at high redshifts. 
Estimates of the dynamical mass of ultra-faint dwarf galaxies (UFDs) within their half-light radius constrain their virialized halo mass before their infall into the Milky Way.
The inferred halo mass and formation redshift of the UFDs place upper bounds on the primordial comoving magnetic field, $B_0$.
We derive an upper limit of $0.50\pm 0.086$ ($0.31\pm 0.04$) nG on $B_0~$ assuming the average formation redshift of the UFD host halos 
is $z_{\rm form}=$ 10 (20), respectively.
\end{abstract}

\section{Introduction}

Magnetic fields in the range between $\sim 10^{-6}$ G to $10^{-3}$G had been observed in local and high redshift galaxies \citep{Bernet:2008gl,Robishaw:2008bq,Fletcher:2011gd,McBride:2013bc,Beck:2015fu,Han:2017jo} 
in different interstellar medium phases.
Limits on the existence of a primordial magnetic field (PMF) of the order of a few nG have been achieved through the data on the Lyman-$\alpha$ forest, Sunyaev-Zel'dovich effect statistics \citep{Pandey:2012kp,Kahniashvili:2012bp}, 
and the cosmic microwave background (CMB) anisotropies \citep{Kosowsky:1996bk,Barrow:1997fr,Shaw:2010ej,Collaboration:2015fj,Zucca:2017jd,Paoletti:2019cg}.
Various theories have been proposed for the generation of magnetic fields on large scales and their amplification \citep{Ichiki:2006cd,Ryu:2008hs,Naoz:2013er}, and on small scales \citep{Widrow:2002ke,Hanayama:2005fb,Safarzadeh:2018df}; however, the origin of the observed magnetic fields remains elusive. 

The presence of seed magnetic fields can affect structure formation. 
The decay of the PMFs through ambipolar diffusion and magnetohydrodynamic turbulence could heat up the intergalactic medium
\citep{Subramanian:1998ba, Sethi:2005dp}. This phenomenon increases the filtering mass, the minimum halo mass that can collapse at a given redshift, which provides the threshold for star formation \citep{Marinacci:2015ee,Marinacci:2016ci}.

The halo mass of old galaxies can be used to set upper limits on the strength of the comoving magnetic field.
The most stringent upper limits can be derived for the smallest collapsed objects in the local universe, namely the ultra-faint dwarf (UFD) galaxies.
UFDs \citep{Brown:2012jo,Frebel:2012ja,Vargas:2013ei} are dark matter dominated galaxies \citep{Simon:2007ee} with total luminosities of $L_{\star}\approx10^3-10^5\lsun$.
The stellar populations of UFDs are all very old \citep[$>12\,$Gyr][]{Brown:2014jn,Weisz:2014cp} implying that UFDs formed most of their stars prior to reionization \citep[e.g.][]{Bullock:2000bn,Bovill:2009hg,Bovill:2011bk}.
 Observations of the population of UFDs are not complete \citep{Tollerud:2008eq}, and their discoveries continues in the Milky Way \citep{DrlicaWagner:2015gb,Koposov:2015cw,Bechtol:2015bd}.

An estimate of the formation redshift and halo mass of the UFDs provides the necessary data to constrain the presence of PMFs at their formation redshift. 
This approach requires two different input data: (i) {\it the stellar population age}; by analyzing the color-magnitude diagram of six ultra-faint dwarfs, \citet{Brown:2014jn} conclude that these systems have formed at least 75\% of their stellar content by $z\sim10$ (13.3 Gyr ago). (ii) {\it the dynamical mass}; such measurements have been carried out for different satellite galaxies of the Milky Way either within different 
3D half-mass radius \citep{Wolf:2010df} or at a fixed physical radius \citep{Strigari:2008ct}.
Combining these data sets and assuming an NFW profile \citep{Navarro:1997if} for the dark matter distribution in these halos, we can infer the halo mass of the Milky Way UFDs at the time of their formation.

The structure of the paper is as follows. In \S (2) we show how to estimate the filtering mass, which is the halo mass scale below which we expect suppressed star formation. 
In \S(3) we use the current data on UFDs to place upper bounds on PMFs' strength, and in \S(4) we summarize our results and discuss their implications.

\section{Filtering mass in the presence of a PMF}

The evolution of baryonic density fluctuations in comoving coordinates in the presence of PMFs follows the equation \citep{Wasserman:1978jo,Sethi:2005dp,Schleicher:2008ba,Sethi:2010jp}, 

\begin{equation}
{\partial^2 \delta_{\rm b} \over \partial t^2} +2 {\dot a \over
a}{\partial \delta_{\rm b} \over \partial t } =4 \pi G \rho _{\rm m}
\delta_{\rm m}  + \delta S(t),
\label{baryon-den-eq}
\end{equation}
where $t$ is the cosmic time, $\delta_m$ is the fractional total matter density perturbation,
\begin{equation}
\delta_m \equiv \frac{\delta \rho_m}{\bar{\rho_m}},
\end{equation}
with similar definition for the baryons ($\delta_b$), and $a=1/(1+z)$ is the cosmological scale factor.
The magnetic source term is
\begin{equation}
S(t)={ \nabla \cdot \left( (\nabla \times {\bf B}) \times {\bf B }\right) \over 4 \pi \rho_{{\rm b}}
a^2 }.
\end{equation}
We write 
\be
{\bf B} = {\bf \bar{B}} + \delta {\bf B}={\bf \bar{B}}+\sum_{\bf k}^{} \delta {\bf B_k}~{\rm exp} (i {\bf k} \cdot {\bf x})
\ee
with ${\bf \bar{B}}$ being the background magnetic related to the comoving magnetic field strength by:
\begin{equation}
\bar{B} a^2 = B_0,
\end{equation}

The magnetic source term can be decomposed into magnetic pressure and magnetic tension terms:
\begin{equation}
 \nabla \cdot [(\nabla \times {\bf B}) \times {\bf B }]= \nabla \cdot[- \nabla(\frac{B^2}{2}) + (\bf{B} \cdot \nabla) \bf{B}]
\end{equation}
which can be written as:
\begin{equation}
- \nabla^2(\frac{B^2}{2})+ \nabla \cdot[(\bf{B} \cdot \nabla) \bf{B}]
\end{equation}

The tension term amounts amount to zero:
\be
{\bf{\bar B}\cdot \nabla}\delta {\bf B} = i \sum_{k} {\bf k\cdot \bar{B}}~\delta{\bf B_k}~{\rm exp}~(i{\bf k\cdot x})
\ee
and therefore,
\be
\nabla \cdot ({\bf{\bar B}\cdot \nabla}\delta {\bf B})= -\sum_{k} ({\bf k\cdot \bar{B}})~({\bf k\cdot} \delta{\bf B})~{\rm exp}~(i{\bf k\cdot x})=0
\ee
because 
\be
\nabla\cdot\delta{\bf B}= 0= i \sum_k {\bf k\cdot} \delta{\bf B}~{\rm exp}~(i{\bf k\cdot x})
\ee
Since the field is real, $\delta {\bf B_{-k}}=\delta {\bf B_{k}^*}$, the pressure term is
\be
\nabla^2\mid {\bf B} \mid^2 =  - \bar{\bf{B}} \cdot \Big[ \sum_{\bf k}^{} k^2 \delta {\bf B_k}~{\rm exp} (i {\bf k} \cdot {\bf x}) \Big].
\ee
Performing the same Fourier decomposition for $\delta_m$, and only keeping the first order terms, 
the leading-order perturbation for the pressure term becomes:
\begin{equation}
\delta S(t)=-\frac{k^2 \bar{B} \delta B}{ 4 \pi \rho_{\rm b} a^2 }.
\end{equation}
Given that $\bar{B} a^2$ is conserved due to flux freezing in ideal MHD assumption, we arrive at $ \bar{B} \propto \rho_b^{2/3}$, which leads to,
\begin{equation}
\frac{\delta B}{\bar{B}} = \frac{2}{3} \frac{\delta \rho_b}{\rho_b}=\frac{2}{3}\delta_m,
\end{equation}
where in the last equality we have assumed baryons perfectly trace dark matter in that the density contrast for both baryons and dark matter are the same on magnetic Jeans scales. It it is unlikely that mechanisms such as streaming velocities could change our results given the formation redshifts considered in our work \citep{Schauer:2019dm}.

We compute the comoving magnetic Jeans wavenumber by setting the right hand side of equation (1) to zero,
\begin{equation}
k_J=\frac{4 \pi a}{\bar{B}} \sqrt{\frac{3}{2}G \rho_b \rho_m}.
\end{equation}

The corresponding physical magnetic Jeans length is,
\begin{equation}
\lambda_J=\frac{2 \pi a}{k_J},
\end{equation}
and the magnetic Jeans mass is given by,
\begin{equation}
M_J^B=\frac{4 \pi}{3} \rho_m (\frac{\lambda_J}{2})^3.
\end{equation}

The filtering mass which is the minimum halo mass that can collapse at a given redshift is computed by the following integration \citep{Gnedin:1997cw}: 
\be
\vspace*{.5cm}
{M_f^B}^{2/3}= \frac{3}{a}\int_0^{a} da^{\prime} {M_J^B}^{2/3} (a^{\prime}) \Big[1-(\frac{a^{\prime}}{a})^{1/2}\Big]
\ee
The result is best fit by,
\be
M_f^B \approx 2\times 10^{8} (B_0/1 \rm nG)^{3} \msun.
\ee 
The redshift evolution of this relationship is negligible at high redshifts. We note that in arriving at this expression we have ignored thermal gas pressure, 
however, the thermal Jeans mass is negligible compared to the magnetic Jeans mass at redshifts relevant to the formation of the UFDs. 
\section{Constraining $B_0$ from observations of the UFDs}

\begin{figure*}
{\vspace*{.5cm}
\includegraphics[width=0.5\textwidth]{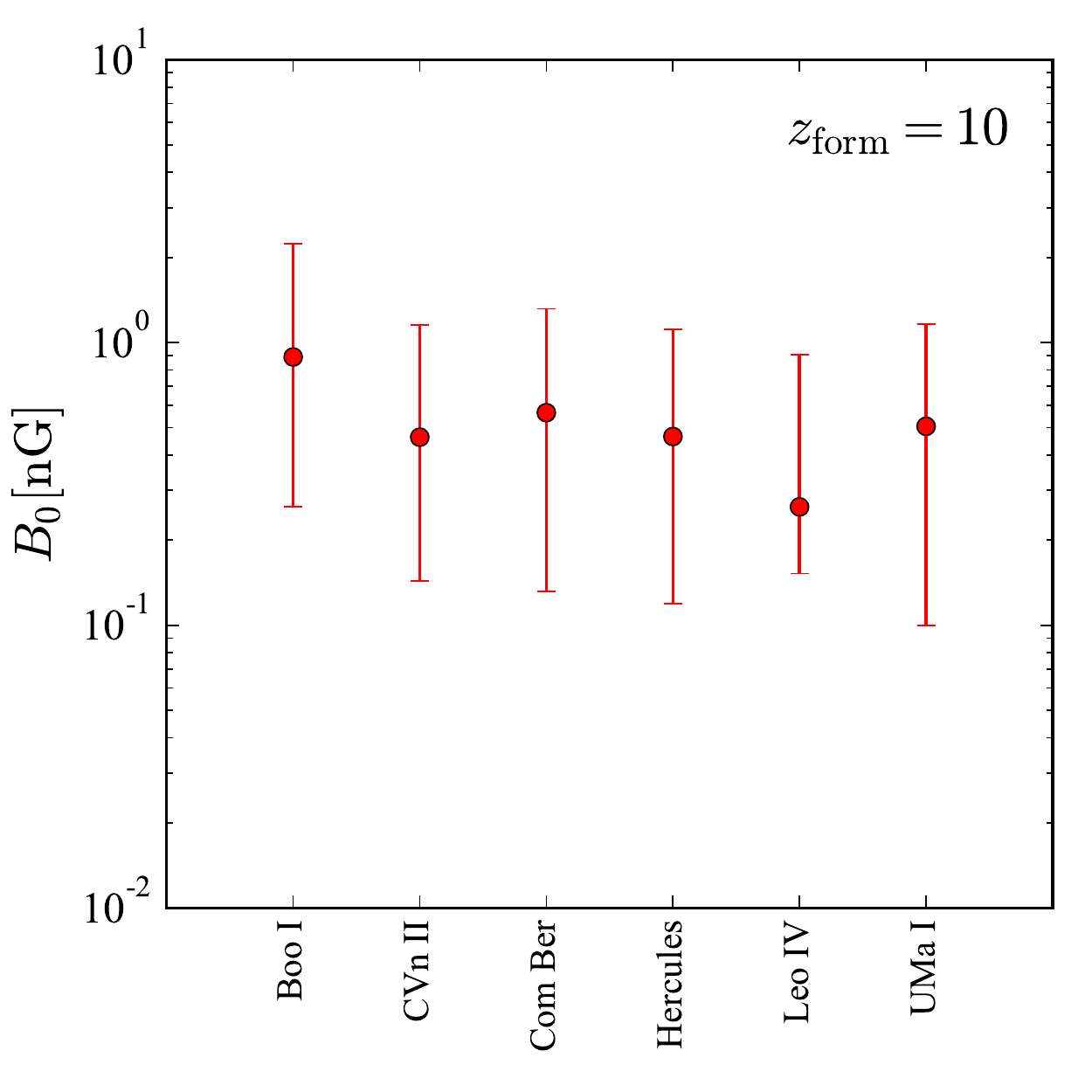}
\includegraphics[width=0.5\textwidth]{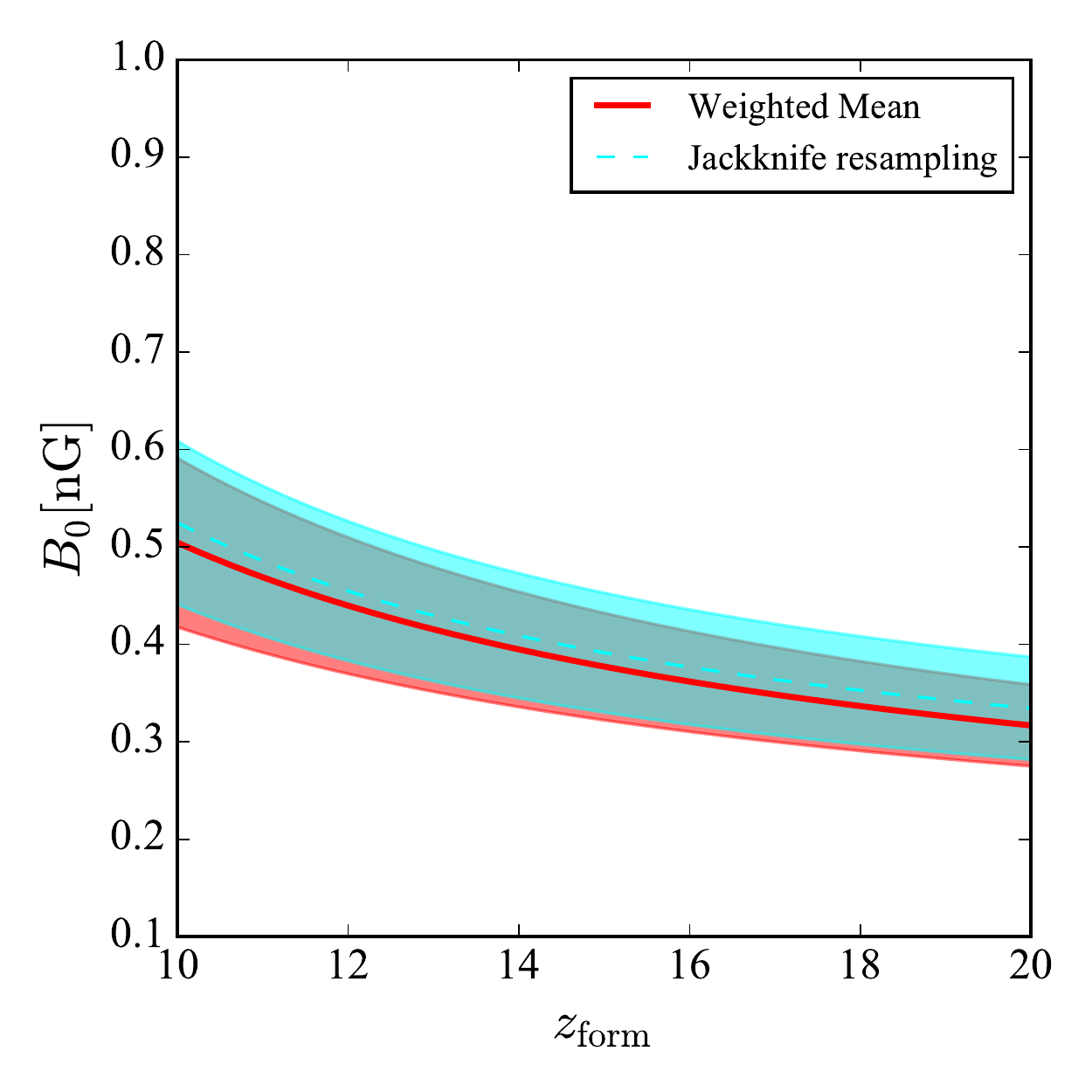}}
\caption{Constraints on the strength of the comoving primordial magnetic field, $B_0$, from observations of UFDs.
{\it Left panel}: the mean upper limit obtained for each of the UFDs assuming a formation redshift of $z_{\rm form}=10$ for all of them. 
The error bars correspond to $1-\sigma$ uncertainty in the $M_{1/2}$ value of each galaxy \citet{Wolf:2010df}. 
We have only considered six UFDs studied in \citet{Brown:2014jn} with known star formation history. {\it Right panel}: the weighted mean upper limit on $B_0$ from combining the results for all six UFDs given an 
assumed formation redshift for their host halos. 
The shaded red region corresponds to the error on the weighted mean. The upper limit on $B_0$ varies from $0.50\pm 0.086 \rm~nG~$ to $0.31\pm 0.04 \rm~nG~$ assuming the formation redshift of their host halo 
to be $z_{\rm form}=$ 10 and 20, respectively. The cyan dashed line and shaded region show the estimate of the mean $B_0$ and its corresponding standard deviation through Jackknife resampling. 
Similarity of the two results implies that our results are not driven in our results by outliers in the sample.} 
\label{f.filter}
\end{figure*}

The enclosed mass within half light radius is related to the observed line of sight velocity dispersion \citep{Wolf:2010df} by,
\be
M_{1/2}\approx \frac{3 <\sigma_{\rm los}^2> r_{1/2}}{G},
\ee
where the brackets indicate a luminosity-weighted average and $r_{1/2}$ is the 3D deprojected half-light radius.

The host halo mass for star formation should not have exceeded the filtering mass at the redshift of star formation for halos 
that host UFDs today. 
We fit NFW profiles to the observed $M_{1/2}$-$r_{1/2}$ of the UFDs and estimate the halo mass at a given formation redshift. 
We only consider six UFDs studied by \citet{Brown:2014jn}:  Bootes I (Boo I), Canes Venatici II (CVn II), Coma Berenices (Com Ber), Hercules, Leo IV, and Ursa Major I (UMa I), where their star formation history has been reliably estimated. 
Since the selected halos have formed most of their stellar mass at redshifts $z>10$, we assume the halos have a concentration parameter of $c=3$ \citep{Correa:2015go}. 
The estimated halo mass is set equal to the filtering mass and the corresponding comoving magnetic field strength associated with that filtering mass is computed. 
The derived halo masses for some of the UFDs fall below the atomic cooling limit at the assumed formation redshifts, however, it is possible for halos with masses below the atomic cooling limit to form stars \citep{Machacek:2001fq,Wise:2007ds}.
The statistics of our derived halo masses is in agreement with the results of high-resolution N-body simulations \citep{Safarzadeh:2018fi}. 

The left panel of Figure 1 shows the results for the six UFDs under consideration. For each galaxy, the mean upper limit on $B_0$ is shown with its $1-\sigma$ error bars corresponding to the uncertainty in the $M_{1/2}$ measurements. 
The errors on $M_{1/2}$  are dominated by observational uncertainties rather than theoretical uncertainties associated with modeling the velocity dispersion anisotropy in these galaxies \citep{Wolf:2010df}.
The mean upper limits on $B_0$ obtained range from  $0.3~ \rm~ nG$ (based on Leo IV) to  $0.9~ \rm~ nG$ (based on Bootes I).

In order to estimate the mean upper limit on $B_0$, we combine the six data points through the weighted mean scheme,
\begin{equation}
\bar{B_0} = \frac{\sum_{i=1}^n{w_i B_{0,i}}}{\sum_{i=1}^n{w_i}}.
\end{equation}
The weights are defined as:
\begin{equation}
w_i=\frac{1}{\sigma_i^2}
\end{equation}
where $\sigma_i$ corresponds to the error bar on the $B_{0,i}$ obtained for each of the UFDs. 
The error on the weighted mean is computed as:
\begin{equation}
\sigma_{\bar{B_0}} = \sqrt{\frac{1}{\sum_{i=1}^n{w_i }}}.
\end{equation}

We show the weighted means for all the six UFDs assuming different formation redshifts for them with a solid red line on the right panel of Figure 1. 
The shaded region corresponds to the weighted mean error for the ensemble of the six UFDs. As expected, assuming higher redshifts of formation for the UFDs 
results in a tighter upper limit on $B_0$. The upper limit on $B_0$ varies from from $0.50\pm 0.086$ ($0.31\pm 0.04$)  nG assuming the formation redshift of their host halo 
is $z_{\rm form}=$ 10 and 20, respectively. 

In order to make sure we are not driven by outliers in our derived upper limits, we redo our analysis through Jackknife resampling, which leaves out one of the 
UFDs each time to estimate the mean upper limit. The result from this method is shown with cyan dashed line and the corresponding standard error with cyan shaded region. 
The two estimates give similar results implying that the outcome is not driven by outliers in the sample.  

\section{Summary and discussion}
We used collapsed dark matter halos at high redshifts to constrain the strength of PMFs.
 The minimum halo mass in which star formation is possible in the presence of a PMF was computed and compared to the 
observed estimates of the enclosed mass within the half-light radius of the UFDs. These galaxies are inferred to have formed the bulk of their stellar mass at redshifts $z>10$, and therefore their host halo mass
should have exceeded the filtering mass at those redshifts. 

Our results based on six UFDs whose star formation histories have been studied in detail imply a stringent upper limit of 
$0.50\pm 0.086$ ($0.31\pm 0.04$) nG for an assumed average formation redshift of their host halo, $z_{\rm form}=$ 10 (20). 
 This limit is better than previously derived limits based on other methods, which range between $1-10~\rm nG~$ \citep{Shaw:2010ej,Pandey:2012kp,Kahniashvili:2012bp,Collaboration:2015fj,Zucca:2017jd,Paoletti:2019cg}, 
 and improve the upper limit of 0.6 nG achieved from CMB non-Gaussianity from the Planck mission data \citep{Trivedi:2014ie}.

In our work we have assumed ideal MHD which is a strong assumption. 
Ideal MHD would indicate coupling of the gas to the PMF, which requires a minimum ionization level to be present at $z>10$
which should be tested against MHD cosmological simulations of structure formation.


\acknowledgements
We are thankful to the referee for the detailed comments which greatly improved our work. 
We are also thankful to Anastasia Fialkov, Josh Simon, and Blakesley Burkhart for helpful comments. 
This work was supported by the National Science Foundation under grant AST14-07835 and by NASA under theory grant NNX15AK82G as well as a JTF grant. 
MTS is grateful to the Harvard-Smithsonian Center for Astrophysics for hospitality during the course of this work. 

\bibliographystyle{apj}


\end{document}